\documentclass{article}
\usepackage{siunitx}
\usepackage{graphicx}	
\usepackage{amsmath}
\usepackage{amssymb}
\usepackage{mathrsfs}
\usepackage{float}
\usepackage{color,soul}
\usepackage[title]{appendix}
\usepackage[a4paper, total={6in, 9in}]{geometry}
\usepackage{authblk}
\usepackage{multicol}
\usepackage{comment}
\usepackage[font={footnotesize}]{caption}
\usepackage{setspace}
\usepackage{cite}

\begin{document}

\title{Impact of DC bias on Weak Optical-Field-Driven Electron Emission in Nano-Vacuum-Gap Detectors}
\author[1]{Marco Turchetti }
\author[1]{Mina R. Bionta}
\author[1]{Yujia Yang}
\author[1,2]{Felix Ritzkowsky}
\author[3]{Denis Ricardo Candido}
\author[3]{Michael Flatt\'e}
\author[1]{Karl K. Berggren}
\author[1]{Phillip D. Keathley \thanks{pdkeat2@mit.edu}}

\affil[1]{Research Laboratory of Electronics, Massachusetts Institute of Technology, Cambridge, MA 02139, USA }
\affil[2]{Deutsches Elektronen Synchrotron (DESY) \& Center for Free-Electron Laser Science, Notkestra\ss e 85, 22607 Hamburg, Germany }
\affil[3]{Dept. Electrical and Computer Engineering, University of Iowa, Iowa City, IA 52242, USA}

\date{\today}
\maketitle

\begin{abstract}
In this work, we investigate multiphoton and optical-field tunneling emission from metallic surfaces with nanoscale vacuum gaps.  Using time-dependent Schr\"{o}dinger equation (TDSE) simulations, we find that the properties of the emitted photocurrent in such systems can be greatly altered by the application of only a few-volt DC bias.  We find that when coupled with expected plasmonic enhancements within the nanometer-scale metallic gaps, the application of this DC bias significantly reduces the threshold for the transition to optical-field-driven tunneling from the metal surface, and could sufficiently enhance the emitted photocurrents, to make it feasible to electronically tag fJ ultrafast pulses at room temperature.  Given the petahertz-scale instantaneous response of the photocurrents, and the low effective capacitance of thin-film nanoantenna devices that enables < 1~fs response time, detectors that exploit this bias-enhanced surface emission from nanoscale vacuum gaps could prove to be useful for communication, petahertz electronics, and ultrafast optical-field-resolved metrology. 
\end{abstract}

%\doublespacing

\section{Introduction}
\label{S:1}

Developing ultrafast photosensitive devices has become increasingly important as these devices are instrumental for the realization of petahertz electronics as well as the further development of ultrafast metrology and information technology\cite{schotz2019,kruger2012,schiffrin2013,yang2019}. Ideally, a device for these applications would be simultaneously fast, sensitive, and capable of operating at room temperature. Photodetectors based on thermal absorption can be both sensitive and broadband, but, since thermal events happen on longer timescales, they have relatively slow response times. Superconducting detectors, on the other hand, are faster (they can achieve few-picosecond response times~\cite{korzh2020}), broadband, and can achieve single-photon sensitivity \cite{goltsman2001}, but they require a cryogenic environment to work, which makes them expensive and difficult to integrate with other systems.  While nanoscale and low-dimensional hot-electron detectors are easier to integrate and can operate at a similar speed (on the scale of 100s of GHz \cite{gosciniak2020}), this is still orders of magnitude slower than recently demonstrated detectors based on multiphoton or optical-field tunneling emission \cite{rybka2016,zimmermann2019, karnetzky2018}. However, while the instantaneous current response from multiphoton and optical-field devices has demonstrated the capability to respond at petahertz-level frequencies, they also require strong fields ($>$ \SI{10}{\volt\per\nano\meter}) and are less sensitive than many alternative photodetection schemes.

In this work, we theoretically investigate the impact of small DC bias voltages on the photoemission properties of surface-enhanced multiphoton and optical-field emission from broadly tunable photodetectors having nanometer-scale vacuum channels~\cite{rybka2016,yang2019, piltan2018}. In particular, we analyze how the additional DC bias can enable entry into the petahertz regime at reduced incident optical field strengths, while simultaneously improving detection efficiency.  With adequate detection efficiency improvements, such detectors could provide a photodetection scheme capable of operating in ambient conditions that is simultaneously sensitive, ultrafast\cite{karnetzky2018}, compact, and easy to incorporate into integrated photonics platforms.  

\section{Biased Nanoantenna Photodetector Concept}

\begin{figure}[h!]
    \centering
    \includegraphics[width = 0.5\linewidth]{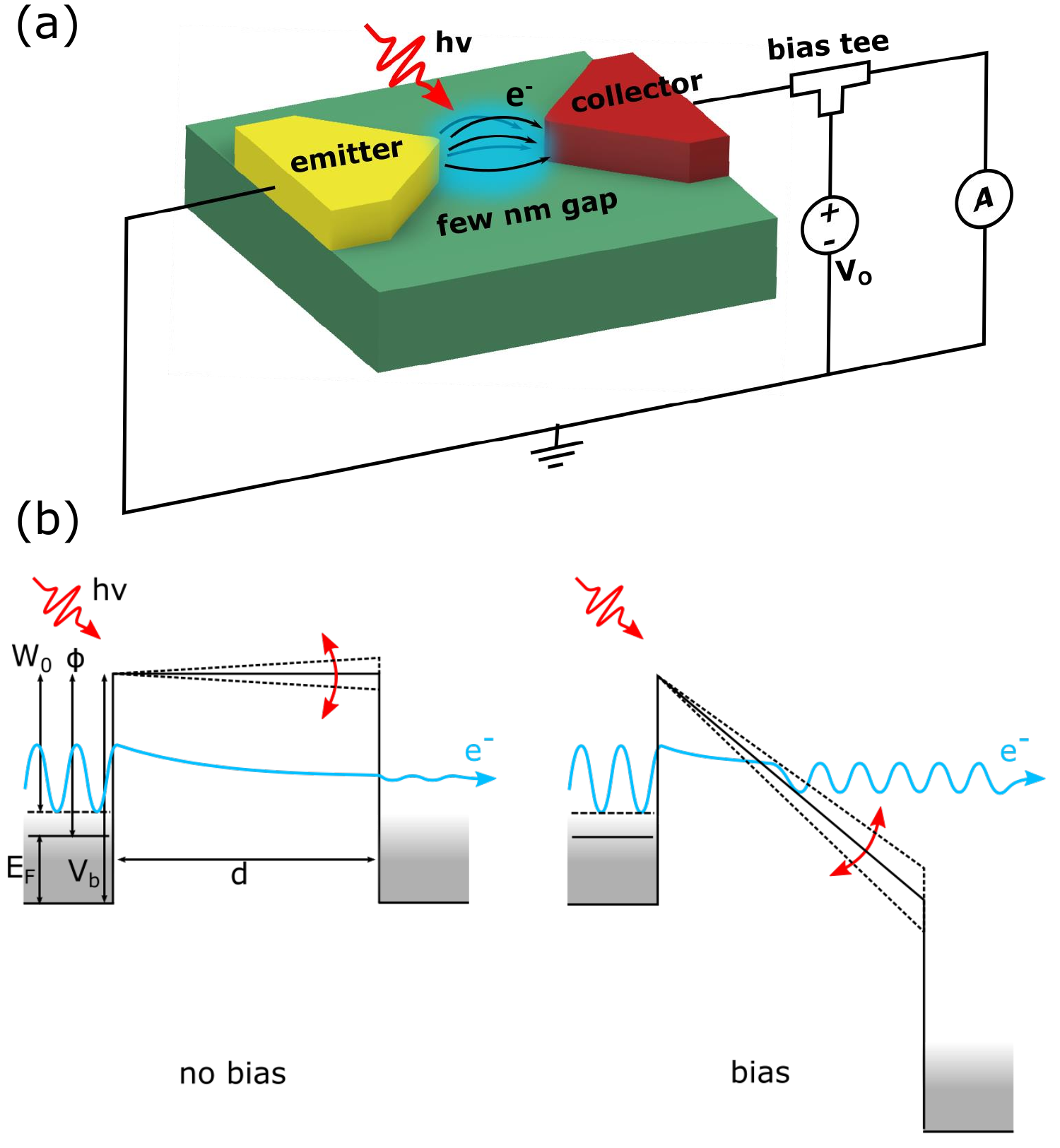}
    \caption{(a) Schematic of the device. A bias is applied in between the emitter (yellow) and the collector (red) of the nanoantenna while an optical pulse impinges on the device perturbing the field emission. (b) Emission mechanism with or without a bias applied between the emitter and collector ($d$ being the gap between the two), showing how the presence of a bias bends the barrier which makes the emission more sensitive to an external optical field.  In the schematic, the Fermi energy $E_F$, the barrier potential $V_b$, the work function $\phi$ and the bound energy state of the emitted electron with respect to the vacuum level $W_o$ are also highlighted. }
    \label{Fig1}
\end{figure}

In Fig. \ref{Fig1}a, we present a schematic of one possible configuration for a nano-vacuum gap optical detector, a plasmonic bowtie nanoantenna sitting on an insulating substrate. The antenna is comprised of an emitter and a collector electrodes, with a nano-vacuum gap on the order of few to tens of nanometers in between (the materials do not necessarily differ and the coloring is simply for visualization). The two electrodes can be comprised in a bowtie configuration where two tips face each other as in the schematic, in a diode configuration where the collector is flat and is adjacent to a tip, or in a slot-waveguide configuration that supports traveling waves that travel within the gap.  The latter case may be preferred in cases where large effective surface areas are desired for increased signal generation.  In all cases, when the gap is very small, the preferential emission direction is dictated by the applied bias\cite{ludwig2020}. The emitter would ideally be fabricated from a plasmonic material, such as gold, enabling large optical field enhancements. For a gold bowtie nanoantenna configuration with an emitter-collector gap of \SI{10}{\nano\meter}, we calculate optical-field enhancements on the order of $100\times$, which is consistent with what has been simulated and observed in similar devices~\cite{piltan2018, yang2019, racz2017}. The emitter and collector are connected to an external readout circuit via a nanowire, that with proper placement does not impact the field enhancement at the metal surface~\cite{yang2019}. When bias voltages of just a few volts are applied to the device, fields on the order of and exceeding \SI{1}{\giga\volt\per\meter} are achieved at the metal surface in the gap region, dramatically modifying the potential barrier of the surface electrons and nearly interacting with the bound electronic states at the surface. 

The potential profile along the length of the device when in a biased and unbiased state is sketched in Fig. \ref{Fig1}b. When unbiased, the bound electronic state on the emitter side has a vanishingly low probability of tunneling through the nano-vacuum gap. The effect of a weak infrared pulse is just that of small modulation of the barrier, which has a limited effect in modifying the emission probability (Fig.~\ref{Fig1}b dashed potential).  Of course, a higher energy ultraviolet photon could excite an electron in the material to an energy state above the barrier for photoemission, but for the sake of this work we are only considering emission from longer wavelengths in the near- to mid-infrared where multiple photons would have to be absorbed to excite the electron to a state above the barrier.  In contrast to the unbiased configuration, under a high bias, the barrier is bent by the applied potential and the tunneling probability increases exponentially.  Thus, by placing a high bias on the collector, even a mild modulation of the barrier due to external alternating optical fields could lead to detectable tunneling photocurrents.  Furthermore, by tuning the structure's size and geometry, the wiggling of the potential at the surface can be greatly amplified due to localized resonances. Together, these two effects make the device highly sensitive to any electromagnetic fields oscillating at the emitter’s resonant frequency.  Unlike conventional photodetection where the photoelectron-conversion probability depends on material band-structures that are difficult to engineer, field-emission depends primarily on the peak surface field strength at the surface and material work function. This distinction means that by tuning the material geometry, it is possible to tune the operating bandwidth of such photodetectors from visible radiation through to the infrared and even into the terahertz range. While the focus of this work is to study the impact of the bias on the photoemission properties we limit our investigation to a single central wavelength in the near infrared (IR). However, this detection scheme could also have a strong impact in the mid-IR\cite{piltan2018} where there is a scarcity of competing technologies.

Due to the high non-linearity of cold-field electron tunneling with respect to the local field outside a material surface, we find that applying a small DC bias between the anode and cathode of the nanoantennas can dramatically increase the probability of emission of a photoelectron from the surface for weak incident optical field strengths (see below where we show a predicted increase of more than 3 orders of magnitude for incident optical field strengths on the order of $\left[ 10^{-4} - 10^{1} \right]$ \SI{}{\volt\per\nano\meter}). Previous studies have investigated the effect of superimposed electric fields from optical and low-frequency electrical sources in different regimes and length scales\cite{Ropers2007,piltan2018,ludwig2020}. However, in this work, we investigate the effect of such a combination of optical and DC bias fields at the nanoscale in the weak-field (sub \SI{10}{\volt\per\nano\meter}) ultrafast (sub \SI{100}{\femto\second}) regime, which has received increasing attention in recent years given the rapid development of nanometer-scale fabrication processes and compact ultrafast optical sources \cite{kruger2018,park2020,schotz2019,ciappina2017}. 

Taking advantage of the nanoscale gaps between the nanoantenna electrodes is crucial for high-speed weak field operation because the nanofabricated sharp tips can provides enhancement of both the DC and optical field and the nanogap allows for ultrafast operation. Subcycle response times of the instantaneous surface current have already been observed with similar structures in Refs. \cite{yang2019, rybka2016, keathley2019, ludwig2020}, and the traveling time of the electrons in the gap is on the order of a femtosecond or less.\cite{karnetzky2018,lee2018}. We calculate that nanoantenna vacuum-gap photodetectors like those used in Ref.~\cite{yang2019} have an effictive capacitance of $\sim$10 aF.  When combined with the few-\si{\ohm}-level resistances of typical antenna structures and connecting wires, this yields a time constant of one unit cell of the device of $RC<\SI{1}{\femto\second}$. We will show that in the few-nm-gap regime, the application of a few-volt DC bias could enable sufficient detection efficiencies to sense fJ-level near-infrared pulses at room temperature. The ability to detect weak optical pulses with such simple, broadband, and high-speed devices would constitute an invaluable contribution to the development of modern ultrafast photodetector technology for applications such as carrier-envelope phase (CEP) detection \cite{yang2019}, lidar\cite{wang2013}, and petahertz-scale optical field sampling \cite{bionta2020}.  

In the following sections, we describe the time-dependent Schrodinger equation (TDSE) simulations used to model the bias-enhanced photoemission process and our simulation results.  In particular, we focus on the impact of high and low static electric fields on the evolution of the wavefunction in the nanoscale gap when excited by ultrafast optical pulses, as well as the dependence of the induced photocurrent with respect to both the optical and static field strengths. Finally, we analyze the impact of the material work function before discussing our results and potential applications.

\section{Bias-Enhanced Photoemission Model}
\label{S:2}
In this section we discuss the model we used to simulate the bias-enhanced photoemission mechanism described in the previous section. First, we approximated the free space photon field as:
\begin{equation}
E_\text{o}(t) = E_\text{o} ~ e^{-\frac{2\log{(2)}t^2}{\tau^2}}\textrm{cos}(\omega t + \phi_{\text{CE}}) \mbox{,}
\end{equation}
where $E_\text{o}$ is the peak field, $\tau$ is the full width half maximum (FWHM) of the pulse, $\omega$ is the frequency and $\phi_{\text{CE}}$ is the phase.
When an optical pulse interacts with the nanoantenna structure, it can excite a surface plasmon which confines the optical field, therefore enhancing it close to each tip (in the gap region between the emitter and collector). 

In the case of a nano-vacuum gap detector, with one electrode at $x=0$ and one at $x=d$, the electric field can be written as:
\begin{equation}
E(x,t) = F E_\text{o}(t) \left( e^{-\frac{x}{L}}+ e^{\frac{x-d}{L}} \right) \mbox{,}
\end{equation}
where $F$ is the field enhancement factor (polarization dependent) at the surface and $L$ is the enhancement decay length, which is typically of the order of \SI{15}{\nano\meter}. 
Therefore the time-dependent potential energy of an electron with charge $q$ is:
\begin{equation}
U(x,t) = \int q E(x,t) \textrm{d}x = - q F E_\text{o}(t) L \left( e^{-\frac{x}{L}}- e^{\frac{x-d}{L}} \right) + C= - q F E_\text{o}(t) 2 L \textrm{sinh} \left( \frac{x-d/2}{L} \right) + C\mbox{.}
\end{equation}
For gaps $d \ll L$ and evaluating the arbitrary constant C so that $U(0,t) =0$, this can be approximated as:
\begin{equation}
U(x,t) \approx - q F' E_\text{o}(t) x \mbox{,}
\end{equation}
where $F'=2 e^{-\frac{d}{2L}}F \approx 2F$ is the field enhancement of the whole nano-vacuum gap detector. Putting this all together, in the biased and illuminated condition we can write the potential energy of the entire structure as
\[    U(x,t)=
                \begin{cases}
                  -q V_\text{b}    & \quad \text{if } x < 0\\
                   + q \left( \frac{V_\text{o}}{d} - F' E_\text{o}(t) \right) x & \quad \text{if } 0 < x < d\\
                  -q V_\text{b} + q V_\text{o} - q F' E_\text{o}(t) d & \quad \text{if } x > d \mbox{,}
                \end{cases}
  \]
where $V_\text{b}$ is the barrier at the interface between the material and vacuum, and $V_\text{o}$ is the DC bias.  

In order to properly account for the total emitted current from the structure, we need to consider emission from the entire ground state electron population.  Since we consider metallic emitters here, we integrated the emission from initial bound energy states $W_\text{o}$, populated according to the Fermi distribution. We also considered the Fermi energy $E_\text{F}$ to have an energy difference with respect to the vacuum level equal to the work function $\phi$. In this framework the kinetic energy of the electrons of a specific bound state is $V_\text{b} - W_\text{o}$. We followed a method similar to that used in Refs.~\cite{fowler_electron_1928, yalunin_strong-field_2011} to estimate cold field emission from static and optical fields, employing a custom TDSE solver, using discrete transparent boundary conditions \cite{Xavier2008}.
Once solved for the wavefunction $\psi$, the transmission coefficient is evaluated as the ratio $\Gamma$ between the transmitted and incoming probability currents, $j_{\text{trans}}$ and $j_{\text{inc}}$ respectively. The following equations are in atomic units.

\begin{equation}
\Gamma (W_\text{o}) = \frac{j_{\text{trans}}}{j_{\text{inc}}} =  \frac{\mathbb{I}\mathrm{m} \left( \psi ^* \frac{\mathrm{d}\psi}{\mathrm{d}x}\right)}{\sqrt{2(V_\text{b} - W_\text{o})}} \mbox{.}
\end{equation}

The total charge density emitted in the time interval $\left[t_0, t_1\right]$ can then be calculated as:
\begin{equation}
    Q = \int_{t_0}^{t_1} \mathrm{d}t \int_{0}^{V_\text{b}} \mathrm{d}W_\text{o} \cdot N(W_\text{o}) \cdot \Gamma (W_\text{o}) \mbox{,}
\end{equation}
where $N(W_\text{o})= k_B T/(2\pi^2) \mathrm{ln} \{ 1+\mathrm{exp} [ (E_F+W_\text{o}-V_\text{b})/(k_B T)]\}$ is the incoming charge supply distribution for each ground state energy.

While the ambient device temperature is considered in determining the initial state electron distribution, we note that we have ignored optically-induced thermal effects.  This is reasonable since prior experimental work has shown that such optically-induced thermal effects play a negligible role in the device emission even for the case of much stronger optical fields when similar nanoantenna emitters are excited with ultrafast optical pulses~\cite{keathley2019, yang2019, putnam_optical-field-controlled_2017, Hommelhoff2006}.  

We also note that we have considered alterations to this simplified potential model due to space-charge reshaping of the barrier and field-penetration into the emitter.  Our photocurrent simulation results when incorporating these effects indicate that they do not impact the primary conclusions of this work, so are not discussed explicitly in this manuscript for the sake of clarity.  In general, the impact of space charge is similar to that experienced by a slight alteration of the effective work function.  Field penetration is rather weak for metals (for example, due to the large magnitude of the material's permittivity, the field strength inside gold is roughly $42\times$ weaker than the surface field at \SI{1}{\micro\meter}), and when incorporated only leads to a slight alteration in the power scaling slope for large optical field strengths.

\section{Simulation Results}
\label{S:3}

We started by simulating a metallic structure with a \SI{5}{\nano\meter} gap, using the model described in the previous section, imposing a work function of \SI{4}{\electronvolt}. We assume Fermi-Dirac statistics in the metallic electrodes and a potential profile as in Fig. \ref{Fig1}b. The fabrication of nanostructures with comparable features has been reported in the literature using a lift-off based patterning processes \cite{keathley2019,rybka2016,zimmermann2019}, where the size of the nanostructures can be adjusted to provide field enhancements on the order of  $100\times$ when a plasmonic material such as gold is employed according to our own electromagnetic modeling. This enhancement is in line with recent experimental observations for similar structures \cite{dombi2013,racz2017}. We perturbed the system with an external optical field waveform assuming a central wavelength of $\lambda = 1\si{\micro\metre}$ and a pulse duration of 30~fs~FWHM. The peak field strength used was \SI{e-2}{\volt\per\nano\meter}. It is worth noting that if the field enhancement of the structure were to be indeed $100\times$, we would obtain this field strength with an applied field of \SI{e-4}{\volt\per\nano\meter}. 

\begin{figure}[h!]
    \centering
    \includegraphics[width = 1\linewidth]{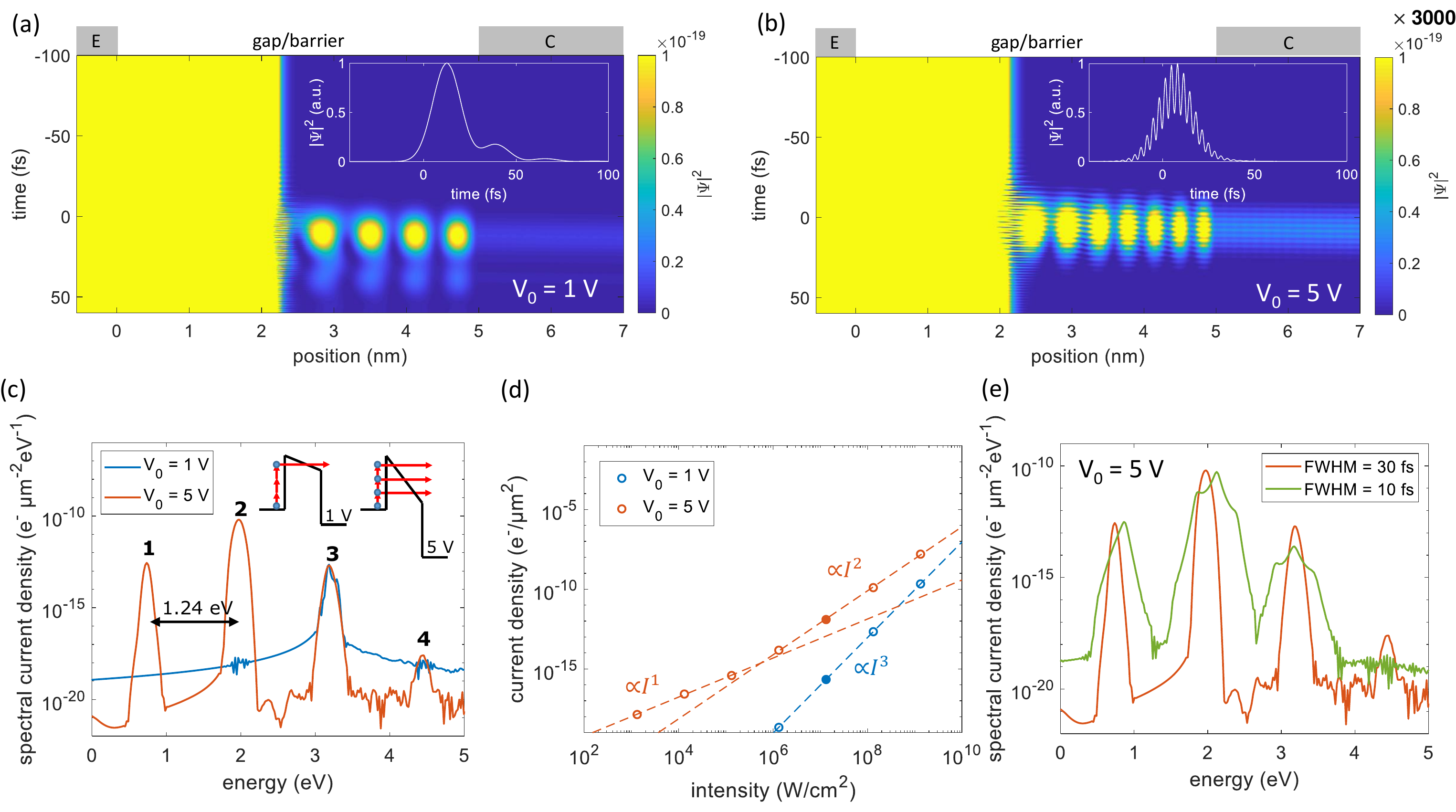}
    \caption{(a-b) Evolution of the right propagating wavefunction inside the structure in a condition of (a) low (1 V) and (b) high (5 V) bias. The optical field used in these simulation was \SI{e-2}{\volt\per\nano\meter}. The position of the emitter, gap/barrier, and collector are also highlighted on top of the graphs. The high bias case exhibits a probability of more than 3 orders of magnitude higher of emitting photoelectrons, as can be seen from the color bar scale. Moreover, in the second case, the fingerprint of the optical pulse is clearly visible in the wavefunction modulation after the barrier. The insets show the wavefunction evolution \SI{1}{\nano\meter} after the barrier. (c) The energy spectrum of the emitted electrons at the two bias conditions. The energy plotted is taken with respect to the initial Fermi level $E_F$ and after removing the bias potential. In the low bias case, only 3-photon processes are possible which result in a single peak in the spectrum. In the high bias case 1-, 2-, and 3-photon processes are all possible. Also the 4-photon process peak starts to appear, which corresponds to the case of an electron that absorbs enough energy to go over the barrier without any tunneling. The 3 peaks are separated by 1.24 eV, consistent with the fact that the optical field is generated by \SI{1}{\micro\meter} photons. The inset shows a cartoon representation of the emission process.  (d) Dependence of the current density on the optical intensity $I$, showing that the low bias case (blue curve) follows a 3rd order power rule ($I^3$), while the high bias case (red curve) follows a 2nd order power rule ($I^2$). This is consistent with the highest peak in (c) being the one corresponding to a 3-photon and 2-photon process in the 1 V and 5 V bias cases respectively. (e) The energy spectrum of the emitted electrons due to optical pulses with a duration of \SI{30}{\femto\second} (red curve) and \SI{10}{\femto\second} (green curve).}
    \label{Fig2}
\end{figure}

In Fig. \ref{Fig2} a,b the result of this simulation is plotted. In particular, we show the wavefunction evolution under the potential barrier inside the gap region for two different bias conditions: \SI{1}{\volt} and \SI{5}{\volt}. This plot shows only the wavefunction of the electrons emitted from the emitter to the collector. The smaller and opposite contribution to the current due to the electrons emitted from the collector to the emitter is also taken into account but not shown here. For the case of a \SI{1}{\volt} bias, the applied bias is not sufficient to bend the barrier enough to significantly enhance the photoemission and the photocurrent signature is very weak. One can also see that a non-negligible amount of the wavefunction is reflected back and forth within the barrier. On the other hand, for a bias of \SI{5}{\volt}, one can clearly see a sub-optical-cycle modulation of the wavefunction and the probability amplitude of the wavefunction is more than 3 orders of magnitude higher than in the 1~V case (the scale of the plot is 3000$\times$ higher).  This sub-cycle modulation is a signature of the optical field enhanced tunneling through the barrier. Therefore, from 1 V to 5 V bias, the system transition from envelope to subcycle emission. Moreover, the amplitude of the tunneled wavefunction in the \SI{5}{\volt} case is more than 3 orders of magnitude higher than for the \SI{1}{\volt} case, as can be clearly seen from the plot scales. 

A semi-periodic wavefunction modulation in time translates to frequency-domain peaks in the wavefunction that are integer multiples of the optical frequency, and thus in the photoelectron energy spectrum we observe photoemission peaks at multiples of the photon energy (the so-called above-threshold photoemission peaks~\cite{schenk_strong-field_2010}) as shown in Fig.~\ref{Fig2}c, where the emitted electron energy is taken with respect to the initial Fermi level $E_F$ and after removing the bias level. In fact for a \SI{1}{\volt} bias only 3 photon transitions are significant, explaining the single photoemission peak.  On the other hand, for the case of the \SI{5}{\volt} bias 1-,2-,3- and 4-photon events are all possible, so we see multiple peaks in the energy spectrum, all spaced by $\hslash \omega \approx \SI{1.24}{\electronvolt}$ apart from each other. This is also demonstrated in Fig.~\ref{Fig2}d when examining the power-law scaling of the photoemission for each biasing condition. Increasing the optical power, or equivalently increasing the peak optical field strength at the nanoantenna surface, increases the probability of extracting an electron. The intensity corresponding to the case portrayed in Fig.~\ref{Fig2}c is highlighted with solid dots. The two curves in Fig.~\ref{Fig2}d show the emission rate as a function of optical power for the two bias cases shown in Fig.~\ref{Fig2}a and b. The \SI{1}{\volt} bias case shows a 3rd power dependence on incident optical power which is consistent with the fact that 3 photon transitions dominate. On the other hand, the \SI{5}{\volt} bias case shows a 2nd power dependence indicating that a 2-photon transition process is the most probable.  This is consistent with our observations in Fig. \ref{Fig2}d where the 2 photon peak is more than one order of magnitude higher than the other photon transition peaks. 
The appearance of multiple peaks when increasing the DC bias suggests the onset of optical-field emission when the Keldysh parameter $\gamma$ is still above unity. This would mean that an external bias can be used to lower the threshold between multiphoton and optical-field emission (or equivalently between weak-field and strong-field) to much lower fields. In fact, while this transition typically happens when $\gamma<1$, which corresponds to $E > \SI{9}{\volt\per\nano\meter}$, with a bias of \SI{5}{\volt} the system appears to be already in optical-field emission regime when $E = \SI{e-2}{\volt\per\nano\meter}$ or $\gamma = 898$.

It is also worth noting that when the pulse gets shorter its bandwidth is larger. This has a direct consequence on the width of the energy peaks. This can be seen in Fig.~\ref{Fig2}e where two pulses with the same peak power but different pulse duration (30 fs and 10 fs FWHM) are analyzed. We can clearly observe that the energy distribution of the electron emitted due to the 10 fs pulse is much wider. This effect could be exploited in a detection scheme that uses energy filtering.

Next, we evaluate how the photocurrent is influenced by the DC bias voltage (or, equivalently, the gap field) in more detail.  

\begin{figure}[h!]
    \centering
    \includegraphics[width = 1\linewidth]{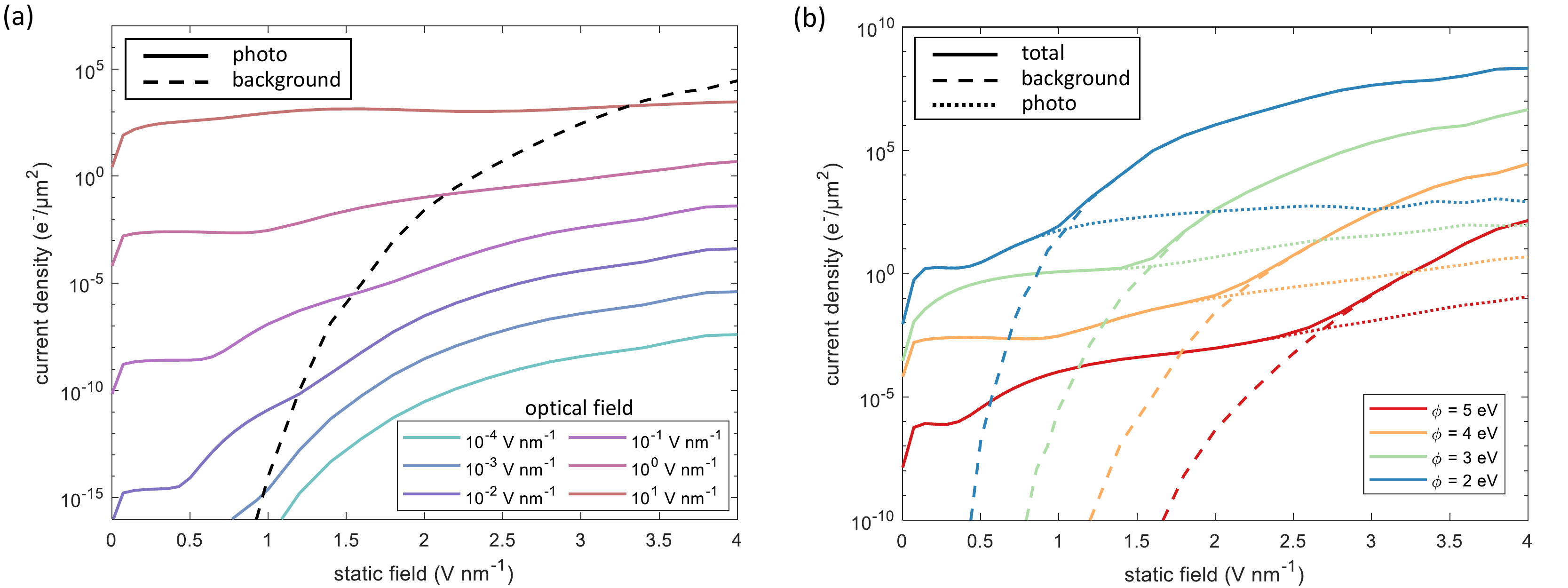}
    \caption{(a) Time-dependent Schrodinger equation (TDSE) simulation of the current density emitted for a $1\si{\micro\metre}$ pulse with 10~fs FWHM pulse duration and optical field ranging from \SI{e-4}{\volt\per\nano\meter} to \SI{e1}{\volt\per\nano\meter}, for different static fields. In this simulation, a work function of $\phi =$ \SI{4}{\electronvolt} was assumed for the emitter and collector material. The total number of electrons extracted for 1 ns time window coming from the background emission due to the bias is shown with a dashed line. (b) TDSE simulation of the current density emitted for materials with work functions $\phi$ ranging from 2 eV to 5 eV extracted in a 1 ns time window with an impinging optical pulse is shown by the solid lines. The fraction of these electrons due to the bias (dashed) and the photon (dotted) are also highlighted. Here the field peak strength was assumed to be \SI{1}{\volt\per\nano\meter}}
    \label{Fig3}
\end{figure}

To differentiate between the photocurrent and current generated from tunneling induced by the DC gap field, we first simulated the current generated in the system in a steady-state condition at different DC gap field levels with no optical field. In particular, we considered static gap field strength between 0 and \SI{4}{\volt\per\nano\meter}, which can be achieved by applying a bias of less than \SI{20}{\volt} using a nanostructure with a \SI{5}{\nano\meter} gap. We then performed the TDSE calculation starting from this initial steady-state with the static gap field applied, incorporating the optical field waveform as a time-dependent perturbation of the potential as shown in Fig.~\ref{Fig1}. To obtain the photocurrent, we subtracted the steady-state current from the total current after activating the optical pulse. We performed the same calculation sweeping the static field magnitude and then repeated the calculation for different optical peak fields with a pulse duration of 10~fs~FWHM,  ranging from \SI{1e-4}{\volt\per\nano\meter} to \SI{1e1}{\volt\per\nano\meter}. Fig. \ref{Fig3}a shows the result of this simulation.

To provide context for the field strengths simulated and how they relate to incident pulse energy, one can think of the bottom curve (with incident optical field $E = $\SI{1e-4}{\volt\per\nano\meter}) in Fig.~\ref{Fig3}a (turquoise) as approximately corresponding to a single photon at \SI{1}{\micro\meter} having a duration of \SI{10}{\femto\second} and focused down to a \SI{1}{\micro\meter\squared} area with field enhancement of 1. This is the worst-case scenario since typical field enhancement values for gold nanoantennas with comparable gap sizes have been shown to range between 20 and 100  \cite{yang2019,dombi2013,racz2017}. The upper limit for peak optical fields is $E = $\SI{10}{\volt\per\nano\meter}, about the level at which the Keldysh parameter $\gamma = 1$ and is borderline between weak and strong field emission.

One possible concern with this detection scheme is the need to differentiate photoelectron emission from the background DC field emission. While the photocurrent increases exponentially with bias voltage, so too does the tunneling current induced by the static field.  The black dashed curve in Fig. \ref{Fig3}a shows simulation results depicting the total number of electrons extracted during a \SI{1}{\nano\second} window, separating the effect of the DC bias and that of the incident optical pulse. This simulation shows that after a certain point the current extracted by the static field becomes much larger than the photocurrent. Therefore, assuming a detector having a \SI{1}{\giga\hertz} rise time, the detection of a single-shot signal should be carried out in a region where the bias provides the maximum possible gain and the photocurrent is still distinguishable from the background.  For instance, for the case of an optical field of $F = \SI{1e-1}{\volt\per\nano\meter}$ in Fig.~\ref{Fig3}a, this operating region would be below \SI{1.5}{\volt\per\nano\meter} of static field strength. A possible workaround that could be used in order to operate the detector at higher DC field levels would be to somehow implement an energy filter on the emitted electrons. As we saw in Fig. \ref{Fig2}c, the photoelectrons have a very specific energy fingerprint compared to those emitted by the static bias field. By only allowing the higher-energy photoelectrons to pass through to the detection electronics, they could be distinguished from those emitted by the static bias. Of course, in an application where the optical signal is repeated at a known frequency, lock-in detection techniques could be used to differentiate the photocurrent from the background DC current allowing for higher static bias fields to be used as well.  However, the electron emission from the static bias will still contribute to the noise floor in this case.  

In Fig.~\ref{Fig3}b, we also study the influence of the material work function on the bias-enhanced photoemission.  Simulations were performed for materials with different work functions $\phi$, ranging from 2 eV to 5 eV, as a function of the gap field.
It is clear from the simulation that a low work function can provide a considerable advantage assuming everything else is equal. However, it should be considered that different materials result in different field enhancements. For example, while gold structures exhibit a larger work function of roughly 5.1 eV, they also provide very large optical field enhancements due to their plasmonic properties.  Materials having much lower work functions can be engineered~\cite{chou2012,koeck2009}, but they are generally unstable and difficult to nanofabricate. Thus, the selection of an optimal material will require a trade off between achievable field enhancements and material work functions.

\begin{figure}[h!]
    \centering
    \includegraphics[width = 1\linewidth]{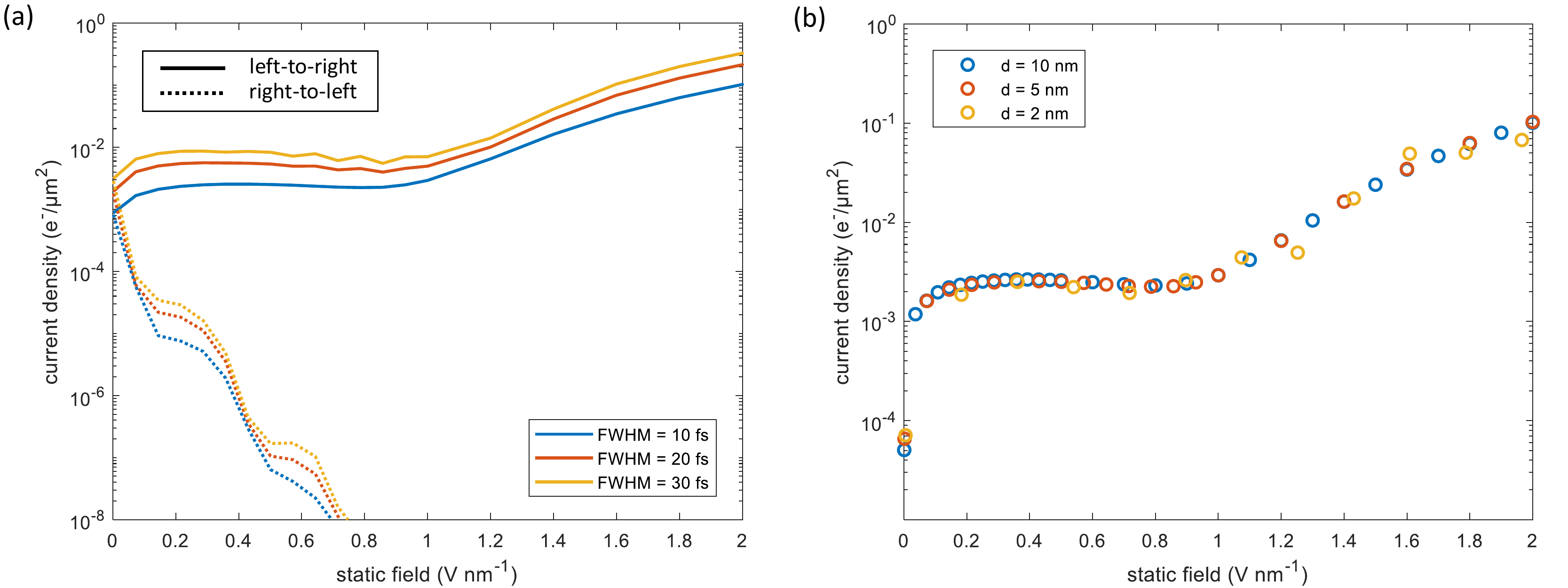}
    \caption{(a) Time-dependent Schrodinger equation (TDSE) simulations of the current density emitted for a $1\si{\micro\metre}$ pulse, a work function of $\phi =$ \SI{4}{\electronvolt}, and a peak optical field strength of \SI{e0}{\volt\per\nano\meter}. The simulation was performed for different pulse duration ranging from $\text{FWHM} =$\SI{10}{\femto\second} to \SI{30}{\femto\second}. The contributions to the photocurrent arising from the emitter to the collector (left-to-right, solid curves) and from the collector to the emitter (right-to-left, dashed curves), are plotted separately. (b). TDSE simulations of the current density emitted for different gap sizes ranging from $d =$\SI{2}{\nano\meter} to $d =$\SI{10}{\nano\meter}. The simulations are performed assuming a pulse with the same characteristics as in (a) and a duration of \SI{10}{\femto\second} .}
    \label{Fig4}
\end{figure}

In Fig.~\ref{Fig4} a and b we investigate the effect on the photocurrent of the ultrafast pulse duration and the gap size, respectively. In both cases, the peak optical field strength was kept constant at \SI{1}{\volt\per\nano\meter}. In the unbiased condition, there is nearly equal amounts of photocurrent excited from the emitter to the collector (left-to-right) and the collector to the emitter (right-to-left), which when combined result in almost zero net photocurrent.  However, since the emission is ruled by a highly nonlinear process as a function of field strength, with sufficient bias the emission in the direction opposite to the preferential one set by the static field is strongly suppressed resulting in a rectified response to the optical field. As shown in Fig.~\ref{Fig4}a, a static field of just \SI{0.1}{\volt\per\nano\meter} is already more than sufficient to achieve almost full rectification. Once rectified, the emitted photocurrent scales almost linearly with the number of optical cycles. We also note in Fig.~\ref{Fig4}b that once the gap size is smaller than the field enhancement decay length so that our approximations hold, it has almost no effect on the emitted current provided the static field is fixed. Of course, we note that obtaining the same bias field with a larger gap would require a larger potential.

\section{Conclusion}
\label{S:2}

We have investigated the impact of a DC voltage bias on photoemission from metallic surfaces surrounding few-nm vacuum gaps.  Nanoscale vacuum gaps enable the application of very large static fields (beyond \SI{1}{\giga\volt\per\meter}) with the application of only a few volts.  Using TDSE simulations, we find that these static fields can have a dramatic impact on the photocurrent induced in the surfaces surrounding the gap.  
In particular, the application of a few-volt bias forced a transition in the system from a condition of envelope driven emission to a sub-cycle response to the external optical field. In fact, the bias appears to lower the threshold between multiphoton and optical-field emission. We showed that optical-field emission may occur in a weak-field condition, with a Keldish parameter of $\gamma = 898$, orders of magnitude higher than the typical requirement of $\gamma \approx 1$ in the case of no additional DC bias.
The bias also causes a dramatic enhancement in the photoemission yield in the weak-field regime, with a typical improvement of more than 3 orders of magnitude in the electron emission probability for incident optical field strengths on the order of $\left[ 10^{-4} - 10^{1} \right]$ \SI{}{\volt\per\nano\meter}.
We also analyzed and discussed how using a material with a lower work function can benefit the sensitivity and the possible trade-offs with the field enhancement.  
The results highlighted by this work can have an important impact on better understanding of the emission mechanisms and in the design of future detectors for multiple applications, from petahertz electronics to lidar to ultrafast metrology.

\bibliography{references}

\section*{Acknowledgements}
This research is supported by the Air Force Office of Scientific Research (AFOSR) grant under contract NO. FA9550-19-1-0065 and by the Army Research Office (ARO) under Cooperative Agreement Number W911NF-16-2-0192.
M.F. also acknowledges support by NSF-DMR1506668.  We thank Prof. William Putnam and Owen Medeiros for their comments and insightful scientific discussion.
\end{document}